\input{aipcheck}

\documentclass[
 ,final            
,numberedheadings     
  ]
  {aipproc}

\layoutstyle{8x11double}

\begin{document}

\title{Statistical Properties of Nuclei by the Shell Model
Monte Carlo Method}

\classification{\texttt{21.60.Cs, 21.60.Ka, 21.10.Ma, 05.30.-d}}

\author{Y. Alhassid}{
  address={Center for Theoretical Physics, Sloane Physics Laboratory,
Yale University, New Haven, CT 06520, U.S.A.}
}

\begin{abstract}
 We use quantum Monte Carlo methods in the framework of the interacting nuclear shell model to calculate the statistical properties of nuclei at finite temperature and/or excitation energies. With this approach we can carry out realistic calculations in much larger configuration spaces than are possible by conventional methods.  A major application of the methods has been the microscopic calculation of nuclear partition functions and level densities, taking into account both correlations and shell effects.  Our results for nuclei in the mass region $A \sim 50 - 70$ are in remarkably good agreement with experimental level densities without any adjustable parameters and are an improvement over empirical formulas. We have recently extended the shell model theory of level statistics to higher temperatures, including
continuum effects. We have also constructed simple statistical models to explain the dependence of the microscopically calculated level densities on good quantum numbers such as parity.  Thermal signatures of pairing correlations are identified through odd-even effects in the heat capacity. 
\end{abstract}

\maketitle


\section{Introduction}

  The statistical properties of nuclei at finite temperature or
excitation energy are important for nuclear astrophysics.  
Level densities are needed for estimates of reaction rates in nucleosynthesis \cite{BBF57}. Such reactions include in particular neutron capture in the $s$ and $r$ processes and proton capture in the $rp$ process. Nuclear partition functions are used in the calculation of thermal stellar thermal reaction rates \cite{rauscher00}.

 Also of interest are the signatures of phase transitions (e.g., pairing) in finite systems. Strictly speaking, phase transitions occur only in the thermodynamic limit (of bulk systems). In finite systems, fluctuations around the mean field are important and they smooth the singularities associated with the phase transitions. An interesting question is whether signatures of these transitions  remain despite the fluctuations. 

 The fundamental quantity for calculating statistical properties
 at temperature $T$ is the partition function
 \begin{equation}
Z(T) =  {\rm Tr} e^{ -H/T} \;,
 \end{equation}
where $H$ is the nuclear many-body Hamiltonian.  Such a Hamiltonian is described by the interacting shell model. The shell model has been successful for describing the properties of, e.g., $sd$-shell and low $fp$-shell nuclei. In such nuclei conventional diagonalization in a complete major shell is possible. However, in medium-mass and heavy nuclei the dimensionality of the  model space is often too large to allow for exact diagonalization.

 Mean-field approximations are tractable but are not
always sufficient. Correlations beyond the mean field can be taken into account by considering fluctuations around the mean field solution.  Formally, this can be expressed by a mathematical transformation known as the Hubbard-Stratonovich (HS) transformation \cite{HS57}. In general, the HS transformation requires an integration over a large number of fluctuating fields.  Quantum Monte Carlo methods have been introduced to integrate over these fields. In the context of the nuclear shell model this approach is known as the shell model Monte Carlo (SMMC) method \cite{LJK93,ADK94,SMMC1}. We briefly review the SMMC method (Section \ref{SMMC}) and then discuss its applications for the calculations of statistical nuclear properties (Sections \ref{level-partition}, \ref{projected} and \ref{pairing}). 

\section{The Monte Carlo Approach}\label{SMMC}

\subsection{Hubbard-Stratonovich transformation}

 The SMMC method is based on the HS transformation \cite{HS57}, which describes the Gibbs ensemble $e^{-\beta H}$  at inverse temperature $\beta$ ($\beta=1/T$)
as a coherent superposition of one-body propagators
$U_\sigma$ 
\begin{equation}\label{HS}
e^{-\beta H} = \int {\cal D}[\sigma]
G_\sigma U_\sigma \;.
\end{equation}
$G_\sigma$  is a Gaussian weight and each $U_\sigma$ describes the imaginary-time propagator of {\em non-interacting} nucleons moving in external time-dependent auxiliary fields $\sigma(\tau)$.
 
 The thermal expectation value of an observable $O$ is given by
\begin{eqnarray}\label{observ}
\langle O \rangle=
{{\rm Tr}\,( O e^{-\beta H})\over{\rm Tr}\,(e^{-\beta H})}=
{\int {\cal D}[\sigma] G_\sigma \langle O \rangle_\sigma{\rm Tr}\,U_\sigma
\over \int {\cal D}[\sigma] G_\sigma {\rm Tr}\,U_\sigma}  \;,
\end{eqnarray}
 where  $\langle O \rangle_\sigma\equiv
 {\rm Tr} \,( O U_\sigma)/ {\rm Tr}\,U_\sigma$ is the expectation value of the observable
$O$ evaluated for a sample $\sigma$ of the auxiliary fields.  

The calculation of the integrands in Eq.~({\ref{observ}) reduces to matrix algebra in the single-particle space. The one-body propagator $U_\sigma$ is represented by an $N_s \times N_s$ matrix ${\bf U}_\sigma$ in the single-particle space ($N_s$ is the number of single-particle orbitals).  The grand-canonical trace of $U_\sigma$ in the many-particle fermionic space can then be calculated from
\begin{equation}\label{partition}
{\rm Tr}\; U_\sigma = \det ( {\bf 1} + {\bf U}_\sigma) \;.
\end{equation}
Eq. (\ref{partition}) describes the grand-canonical partition function of non-interacting fermions in time-dependent external fields $\sigma(\tau)$.  Similarly, $\langle O \rangle_\sigma$ can be expressed in terms of the matrix ${\bf U}_\sigma$ using Wick's theorem.

  In finite nuclei it is important to consider a definite number of protons and neutrons. Therefore the traces in Eq.~(\ref{observ}) should be evaluated in the canonical ensemble.  Canonical traces can be calculated by exact particle-number projection.

   At sufficiently high temperatures, only static configurations of the fields $\sigma$ are important, leading to the static path approximation (SPA) \cite{SPA}. Of particular importance are large-amplitude fluctuations of the relevant order parameters. For example, in the Landau theory of shape transitions, static fluctuations of the quadrupole deformations have explained the observed temperature and spin dependence of the giant dipole resonance \cite{GDR}. 

At low temperatures, it is necessary to take into account all fluctuations including the quantal fluctuations which are described by time-dependent configurations of the fields $\sigma$. This requires an integration over a very large number of variables and in practice can be done by Monte Carlo methods.

\subsection{Monte Carlo methods}

 The multi-dimensional integral is evaluated exactly by Monte
Carlo methods, in which the $\sigma$ fields are sampled according to the distribution 
\begin{equation}
W_\sigma\equiv   G_\sigma \vert {\rm Tr} U_\sigma \vert \;.
\end{equation}
The observables are then estimated from
\begin{equation}
\langle O \rangle \approx { \sum_\sigma \langle O \rangle_\sigma \Phi_\sigma \over \sum_\sigma \Phi_\sigma}
\end{equation}
where $\Phi_\sigma\equiv {\rm Tr} \; U_\sigma /\vert {\rm Tr}\; U_\sigma \vert$
is the sign of the one-body partition function ${\rm Tr}\, U_\sigma$

 Often the sign $\Phi_\sigma$ is not positive for some of the samples $\sigma$. When the statistical uncertainty of the sign is larger than its average value, the method fails. This known as the Monte Carlo sign problem.  Most effective nuclear interactions suffer from this sign problem at low temperatures. A practical solution to this sign problem is discussed in Ref.~\cite{ADK94}. Furthermore, ``good-sign'' interactions can be constructed for realistic calculations of collective properties. Such interactions are used in the applications described below.

 In general, the computational properties of the Monte Carlo approach scale more favorably as a function of the number of single-particle orbitals $N_s$, enabling calculations in much larger configuration spaces than are possible with conventional methods (the Monte Carlo method scales as $N_s^4$ compared with an exponential scaling of a direct diagonalization). 

\section{Nuclear level densities and Partition  functions}\label{level-partition}

The Fermi gas model ignores important correlations. In practice, correlations are taken into account through empirical modifications of the Fermi gas model. In particular, good fits to the data are obtained using the backshifted Bethe
formula (BBF) \cite{gc65}
\begin{equation}\label{BBF}
\rho(E_x) =
 {{\sqrt\pi}\over{12}} a^{-1/4} (E_x - \Delta)^{-5/4} e^{2\sqrt{a
(E_x - \Delta) }} \;,
\end{equation}
where $a$ is the single-particle level
density parameter and $\Delta$ is the backshift parameter.
 However, $a$ and $\Delta$ have to be adjusted for each nucleus, 
and it is therefore difficult to predict the level density to an accuracy better than an order of magnitude.

The interacting shell model includes both shell effects and residual interactions and is therefore a good microscopic model for the calculations of level densities. 
 Truncations suitable for the description of low-lying states cannot be used at finite temperature and thus complete major shells must be included. This requires calculations in large model spaces and SMMC is a suitable approach. In Ref.~\cite{NA97} we have introduced a method to calculate level densities in the SMMC method.

 \subsection{Thermodynamic approach}\label{thermodynamics}
 
The level density can be obtained as the inverse Laplace transform of the partition function. The {\em average} level density is found when this transform is evaluated in the saddle point approximation
\begin{equation}\label{level}
\rho(E) \approx {1 \over \sqrt{2\pi T^2
C}} e^{S(E)} \;,
\end{equation}
where $S(E)$ is the canonical entropy and $C$ is the canonical heat capacity.

 In the Monte Carlo approach, we calculate the thermal energy from $E(\beta) \equiv \langle
H\rangle$ and integrate the thermodynamic relation 
 $ - \partial \ln  Z / \partial \beta = E(\beta)$ to find
 the partition function $Z(\beta)$.
The entropy and heat capacity in (\ref{level}) are calculated from $S(E) = \ln Z + \beta E$ and
  $C = -\beta^2 \partial  E /\partial \beta$, respectively.

\subsection{Level densities in the iron region}\label{iron}

  We have applied the SMMC approach to calculate level densities in the mass region $A \sim 50 - 70$ \cite{NA97,ALN99}. The shell model space includes the complete $fpg_{9/2}$ shell. We have constructed a good-sign interaction that includes attractive monopole pairing and multipole-multipole interactions.  The multipole interaction terms are based on a surface-peaked interaction whose strength is determined self-consistently. The quadrupole, octupole and hexadecupole terms are then renormalized by factors of 2, 1.5 and 1, respectively. The pairing strength is determined from odd-even mass differences.

 We have found remarkably good agreement with experimental level densities without any adjustable parameters. An example is given in Fig.~\ref{fig1}, which shows the level densities of several iron isotopes versus excitation energy. The SMMC level densities (symbols) are compared with the experimental level densities (solid lines).  

\begin{figure}\label{fig1}
  \includegraphics[height=.27\textheight]{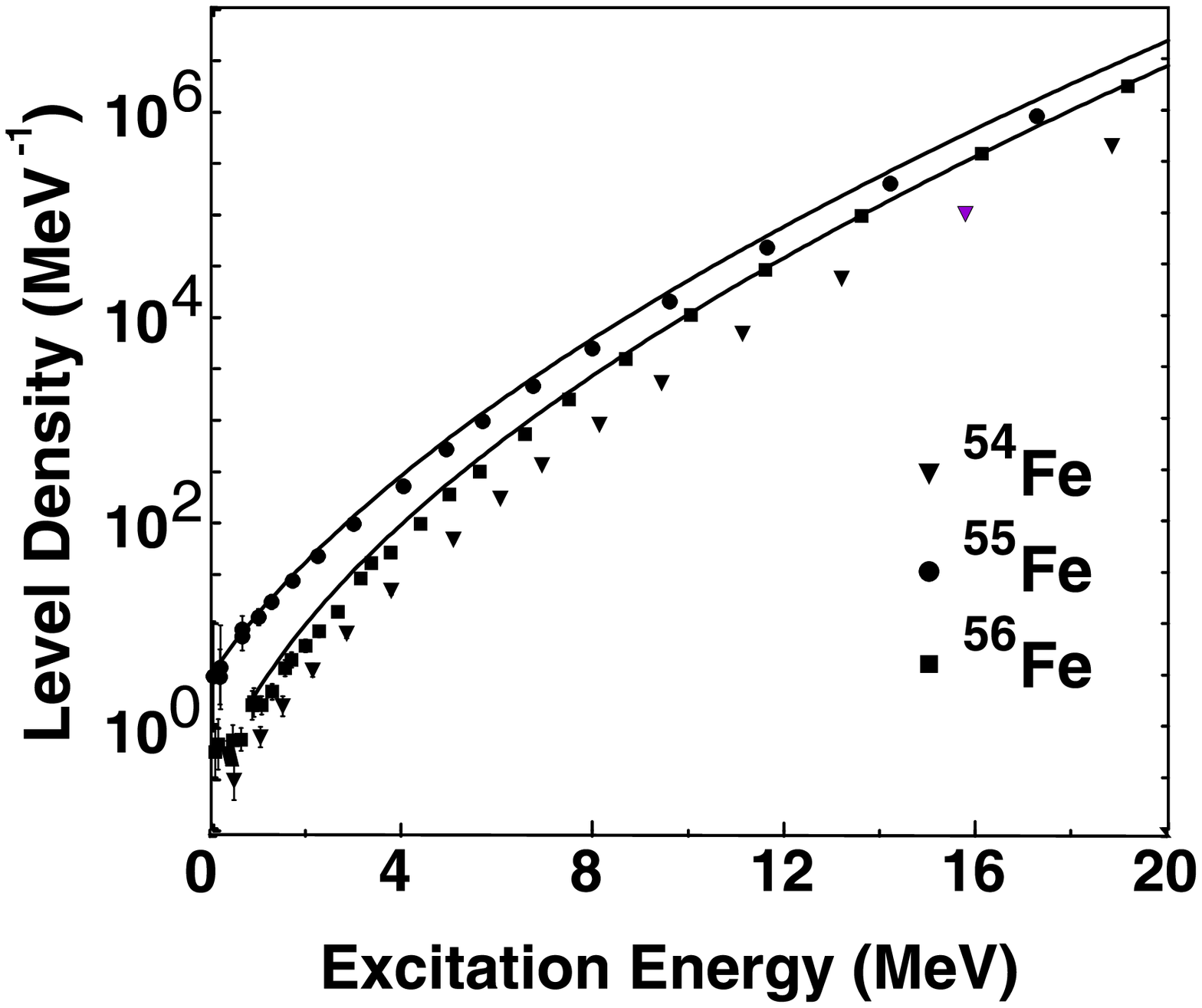}
  \caption{The SMMC level densities of three iron isotopes (symbols). Also shown are the experimental level densities of $^{55}$Fe and $^{56}$Fe (solid lines).}
\end{figure}

The microscopically calculated level densities are well described by the BBF (\ref{BBF}). By fitting the SMMC level densities to the BBF, we can extract the parameters $a$ and $\Delta$. In Fig.~\ref{fig2} we show the dependence of these extracted parameters (solid squares) on mass number $A$ for manganese, iron and cobalt isotopes and compare them with experimental values (x's) \cite{Dilg}. Our microscopic results are an improvement over empirical formulas (solid lines) \cite{HWFZ}. The parameter $a$ varies smoothly as a function of mass while $\Delta$ exhibits odd-even staggering because of pairing effects.

\begin{figure}\label{fig2}
  \includegraphics[height=.3\textheight]{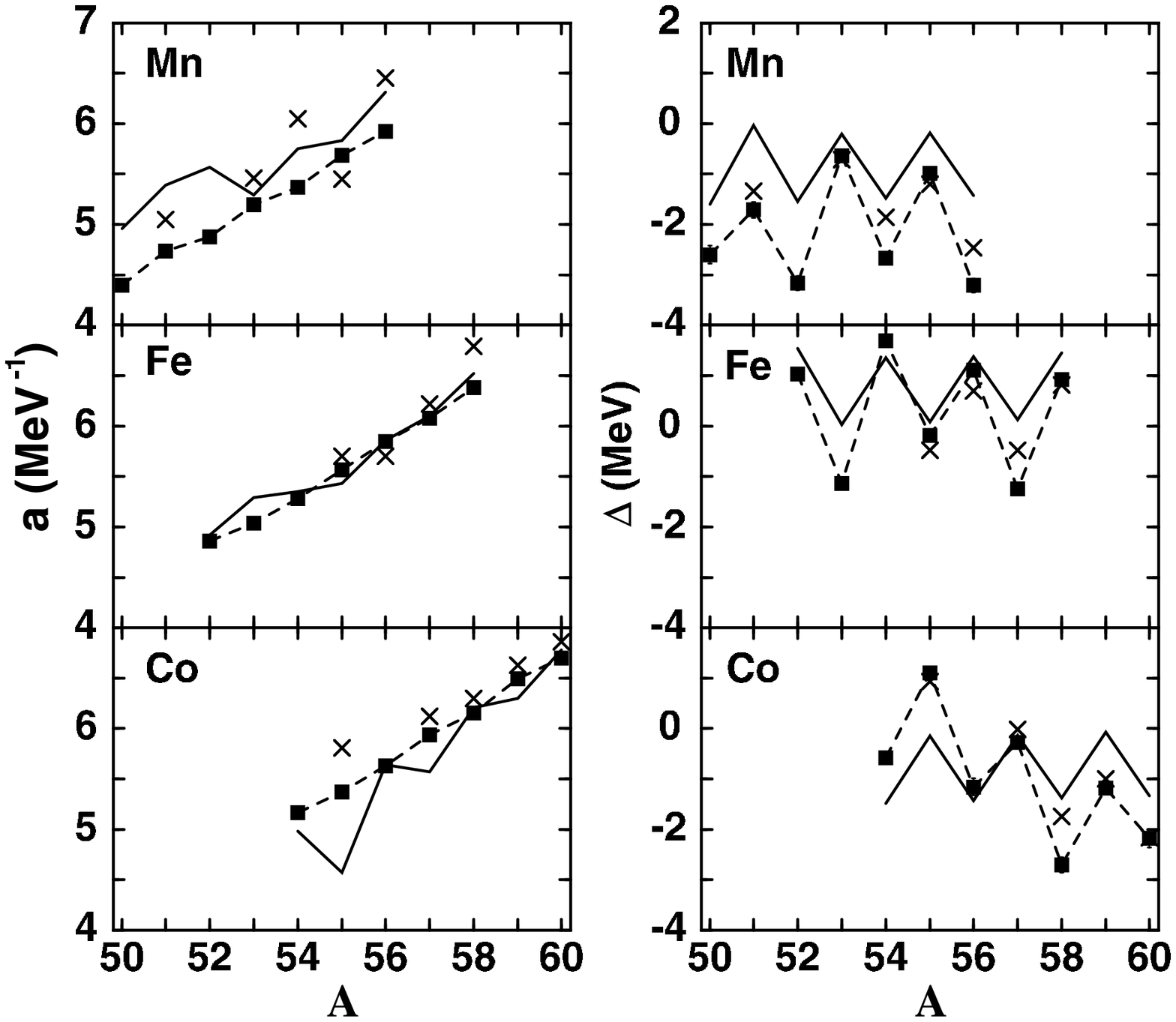}
  \caption{The single-particle level density parameter $a$ (left panels) and the backshift parameter $\Delta$ (right panels) versus mass number $A$ for different isotopes. The SMMC results (solid squares) are compared with experimental values (x's) \cite{Dilg} and empirical formulas (solid lines) \cite{HWFZ}. From Ref.~\cite{ALN99}}
\end{figure}

\subsection{Extending the theory to higher temperatures}

The SMMC calculations of Section \ref{iron} are realistic for 
 temperatures below $\sim 1.5 - 2$ MeV. At higher temperatures one must treat larger model spaces. Extending SMMC to larger spaces is however computationally time-consuming, and we have recently developed an approximate method to extend the theory to higher temperatures while using SMMC only in the truncated space \cite{ABF03}.

In the independent-particle model, it is possible to treat the complete space including all single-particle bound states as well as the continuum. In particular, the many-particle grand canonical partition function $Z^{\rm GC}_{sp}$  can be expressed in terms of the single-particle energies $\epsilon_{n l j}$ and
scattering phase shifts $\delta_{lj}(\epsilon)$
\begin{eqnarray}
\ln Z^{\rm GC}_{sp} & = & \sum_{lj}(2j+1) \left\{\sum_n
\ln [1 + e^{-\beta(\epsilon_{nlj}-\mu)}] \right. \\ \nonumber
&& \left. + 
\int_0^\infty d\epsilon \delta\rho(\epsilon)\ln[1 +
e^{-\beta(\epsilon-\mu)}]\right\}
\end{eqnarray}
where $\delta\rho(\epsilon) =  \pi^{-1}\sum_{l j} (2j+1)
{d\delta_{lj}/d\epsilon}$ is the continuum contribution to the single-particle level density. In the calculations we have used a Woods-Saxon potential with the parametrization of Ref. \cite{BM69} 

  The canonical partition function
 $Z_N$ at fixed particle-number $N$ can be obtained in the saddle-point approximation 
$\ln Z_N \approx \ln Z^{\rm GC} -\beta\mu N -{1\over 2}\ln 
\left( 2 \pi  \langle(\Delta N)^2\rangle \right)$,
where $\langle(\Delta N)^2\rangle$ is the variance of
the particle number fluctuation. 

 In the presence of interactions, we combine the fully correlated SMMC partition in the truncated space with the independent-particle model partition in the full space through
\begin{equation}\label{extended-Z}
 \ln Z_v = \ln Z_{v,tr} + \ln Z_{sp} -\ln Z_{sp,tr} \;.
\end{equation}
Here $Z_v$ and $Z_{v,tr}$ are the partition functions in the presence of interactions in the full and truncated spaces, respectively. The subtraction of the last term on the r.h.s. of Eq.~(\ref{extended-Z}) avoids a double counting of the truncated degrees of freedom. The partition function (\ref{extended-Z}) includes correlations and should be realistic up to $T \sim 4$ MeV (at higher temperatures it is necessary to take into account the temperature dependence of the mean field).

 The logarithm of the excitation partition function $Z' \equiv Z e^{\beta E_0}$ ($E_0$ is the ground-state energy) is shown in Fig.~\ref{fig3} as a function of temperature. The extended partition $Z_v$ (solid squares) is compared with the SMMC partition in the truncated space $Z_{v,tr}$ (open squares) and the independent-particle model partition $Z_{sp}$ (dashed line). The solid line is a fit of the extended partition function to the partition function associated with the BBF \cite{ABF03}. 

\begin{figure}\label{fig3}
  \includegraphics[height=.25\textheight]{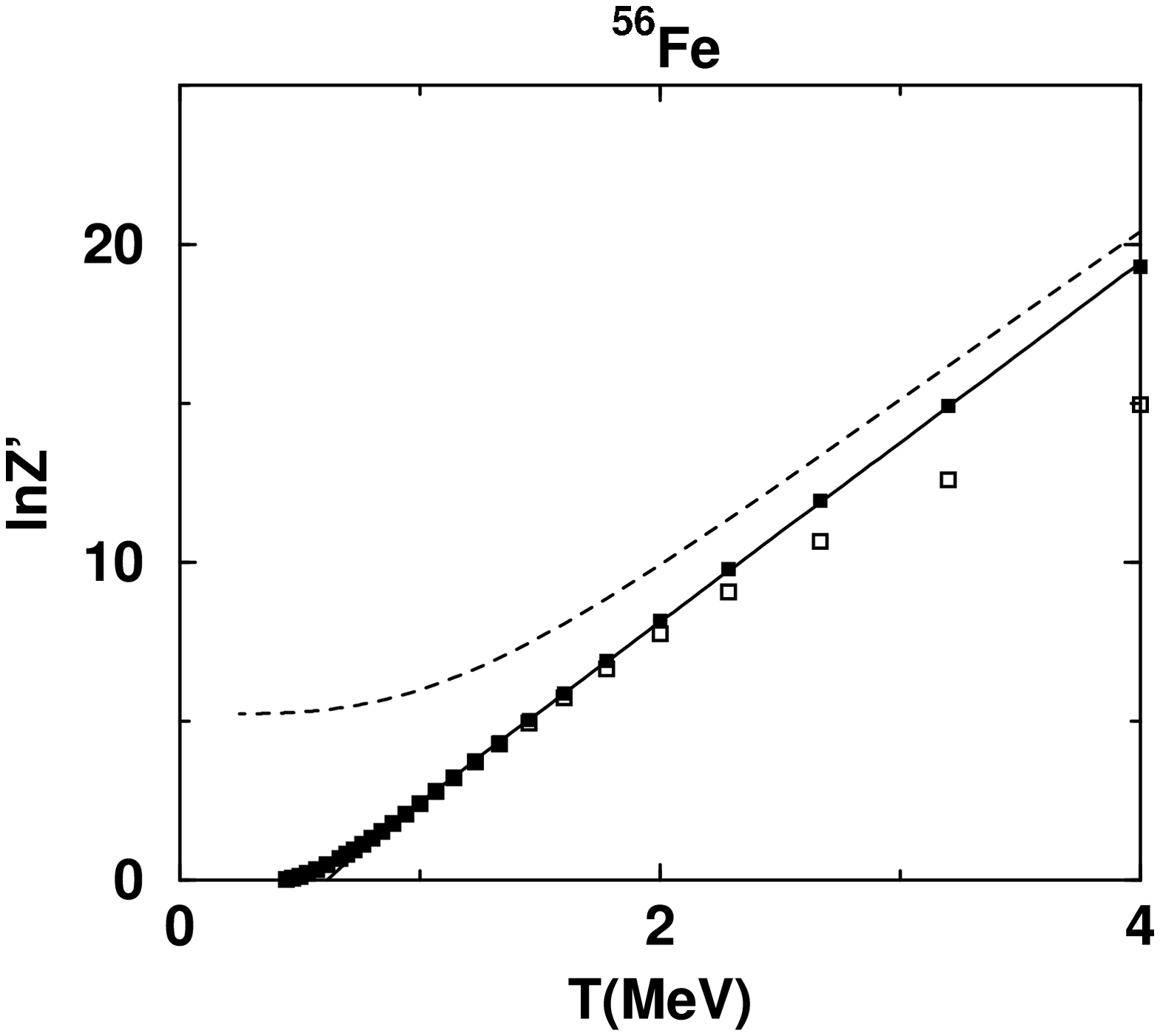}
  \caption{The partition function $Z'$ as a function of temperature for $^{56}$Fe. The SMMC partition in the truncated space (open squares) is combined with the independent particle model partition (dashed line) to give the extended partition function (solid squares). The solid line is a fit to a partition function associated with the BBF. From Ref.~\cite{ABF03}.}
\end{figure}

 The corresponding extended level density for $^{56}$Fe is shown in Fig. \ref{fig4} (solid squares) and is compared with the level density in the truncated space (open squares).  The extended level density is well described by the BBF (solid line) with fixed $a$ and $\Delta$ up to $T \sim 4$ MeV.

\begin{figure}\label{fig4}
  \includegraphics[height=.25\textheight]{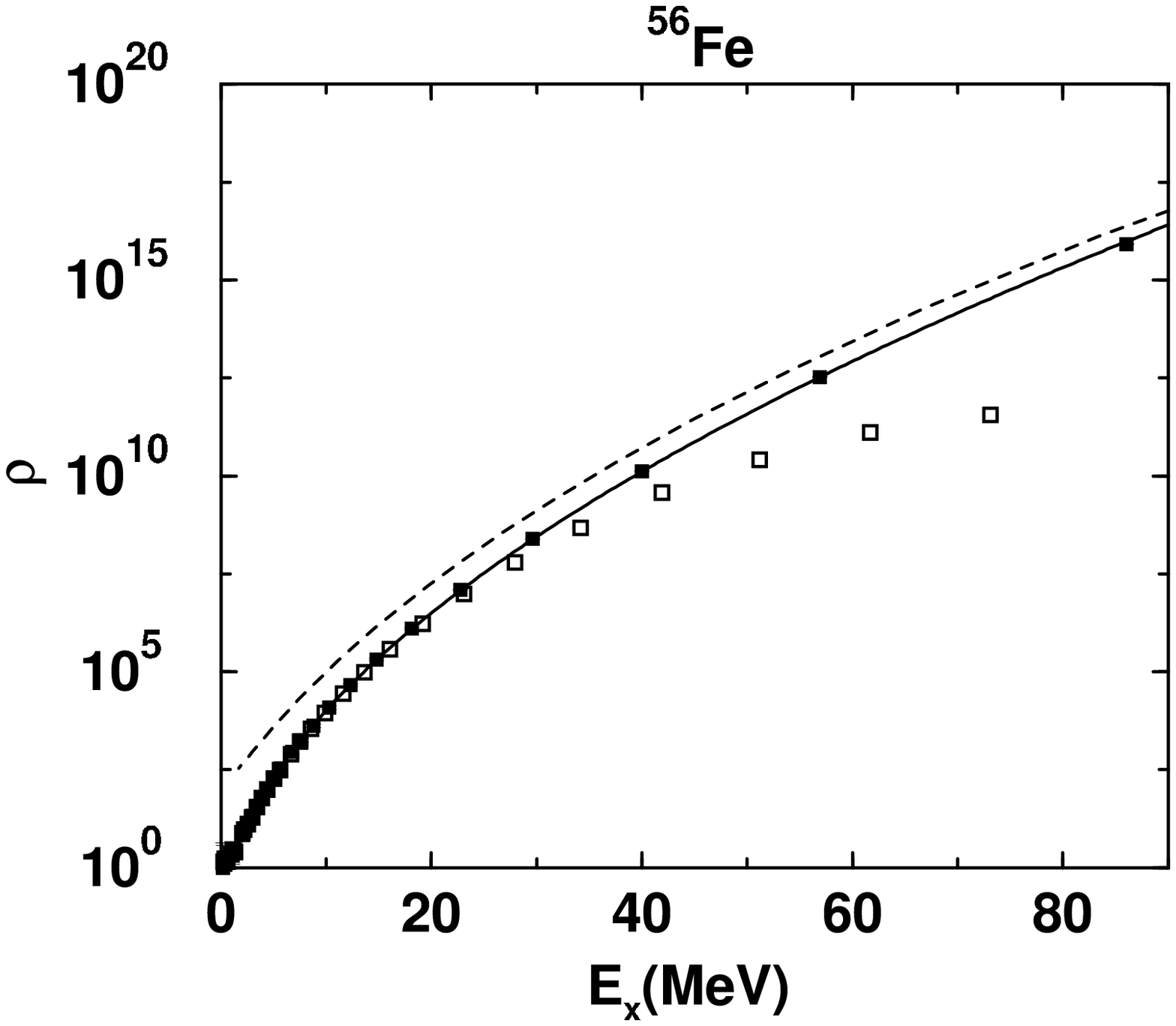}
  \caption{The extended level density of $^{56}$Fe (solid squares) as a function of excitation energy is compared with the SMMC level density in the truncated $fpg_{9/2}$ shell (open squares). The solid line is a fit of the extended level density to the BBF. From Ref.~\cite{ABF03}}.
\end{figure}

\section{Projected level densities}\label{projected}

 It is often necessary to know the dependence of the level density and partition function on the good quantum numbers such as parity, spin and isospin.  In SMMC this can be achieved by introducing the appropriate projections in the HS transformation.

\subsection{Parity distribution}

 We have calculated (in SMMC) the projected energies $E_\pm$ for even- or odd-parity states as a function of $\beta$ and then applied the method of Section \ref{thermodynamics} to find the even- and odd-parity level densities $\rho_\pm$ \cite{NA97,parity}.

Contrary to the assumption often used in nucleosynthesis calculations, we find that in some nuclei $\rho_+\neq \rho_-$  even at the neutron resonance energy.   In Fig.~\ref{fig5} we show the SMMC ratio $\rho_-/\rho_+$ (symbols) versus excitation energy for three nuclei in the mass region $A \sim 50 - 70$. The crossover from one dominating parity at low excitations to equally likely parities at higher energies  depends on the particular nucleus. We have introduced a simple statistical model \cite{parity} to estimate the odd-to-even parity ratio. The model assumes that the quasi-particle states with parity $\pi$ ($\pi$ being the parity with the smaller occupation) are randomly populated. We find for
the ratio of odd- to even-parity partition functions 
${Z_-(\beta)/ Z_+(\beta)} =\tanh f$, where
$f$ is the occupation of quasi-particle states with parity $\pi$. To mimic effects of the quadrupole-quadrupole interaction, we use deformed quasi-particle states. 

  The results of our model (solid lines in Fig.~\ref{fig5}) are in good agreement with the SMMC results (symbols).

\begin{figure}\label{fig5}
\includegraphics[height=0.3 \textheight]{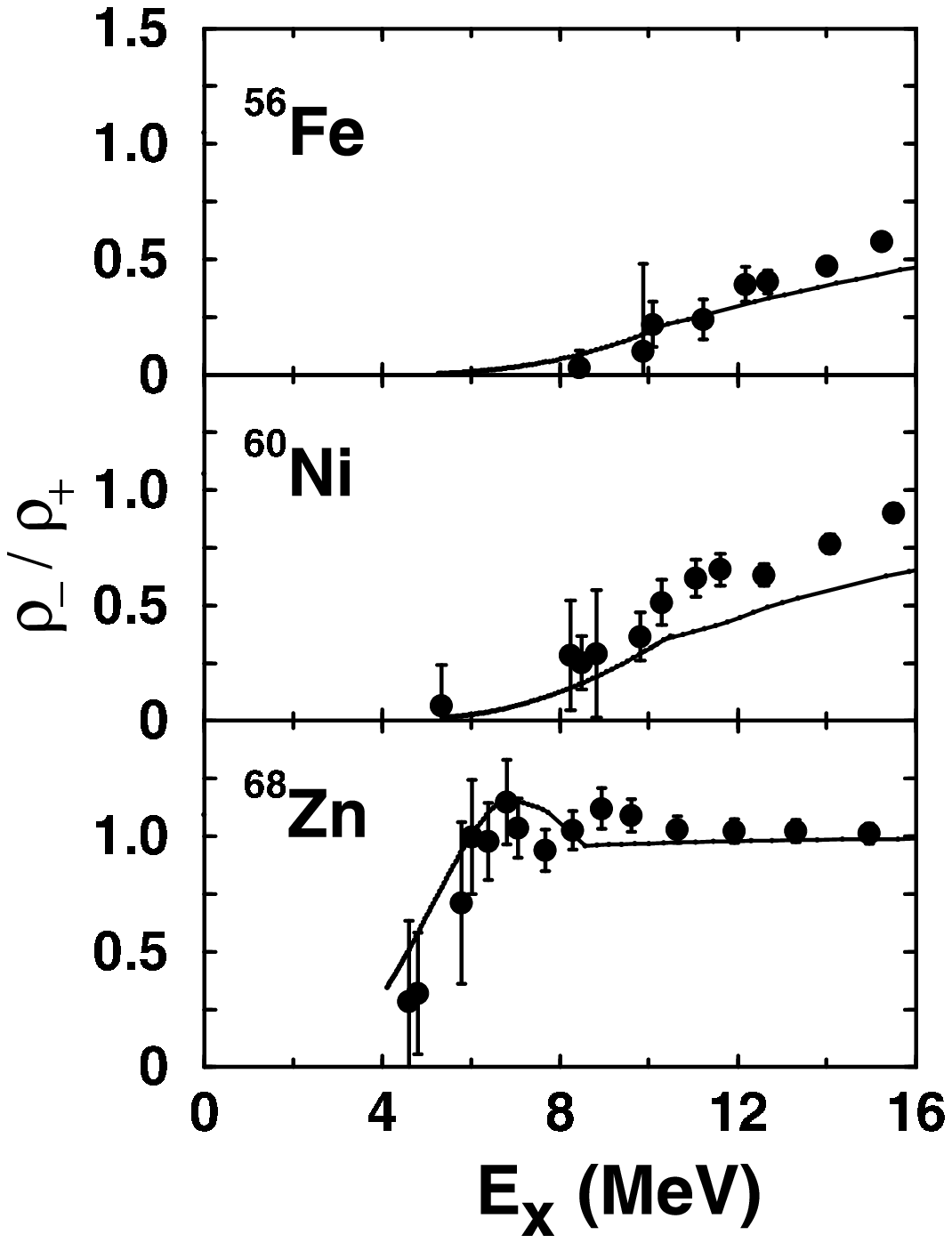}
\caption{The ratio $\rho_-/\rho_+$ of odd- to even-parity level densities versus excitation energy $E_x$ for $^{56}$Fe, $^{60}$Ni and $^{68}$Zn. The microscopic SMMC results are compared with the results of a simple statistical model (solid lines). From Ref.~\cite{parity}}.
\end{figure}

\subsection{Isospin distribution}

 Isospin is approximately a good quantum number in nuclei. The isospin dependence of level densities can be calculated in SMMC by exact isospin projection \cite{NA03}.  Isospin projection also allows us to take into account the proper isospin dependence of the nuclear interaction. This isospin dependence of the nuclear interaction can lead to significant corrections in the total level density of $N \sim Z$ nuclei.

\section{Heat capacity and the pairing transition}\label{pairing}

 \begin{figure}\label{fig7}
\includegraphics[height=0.36\textheight]{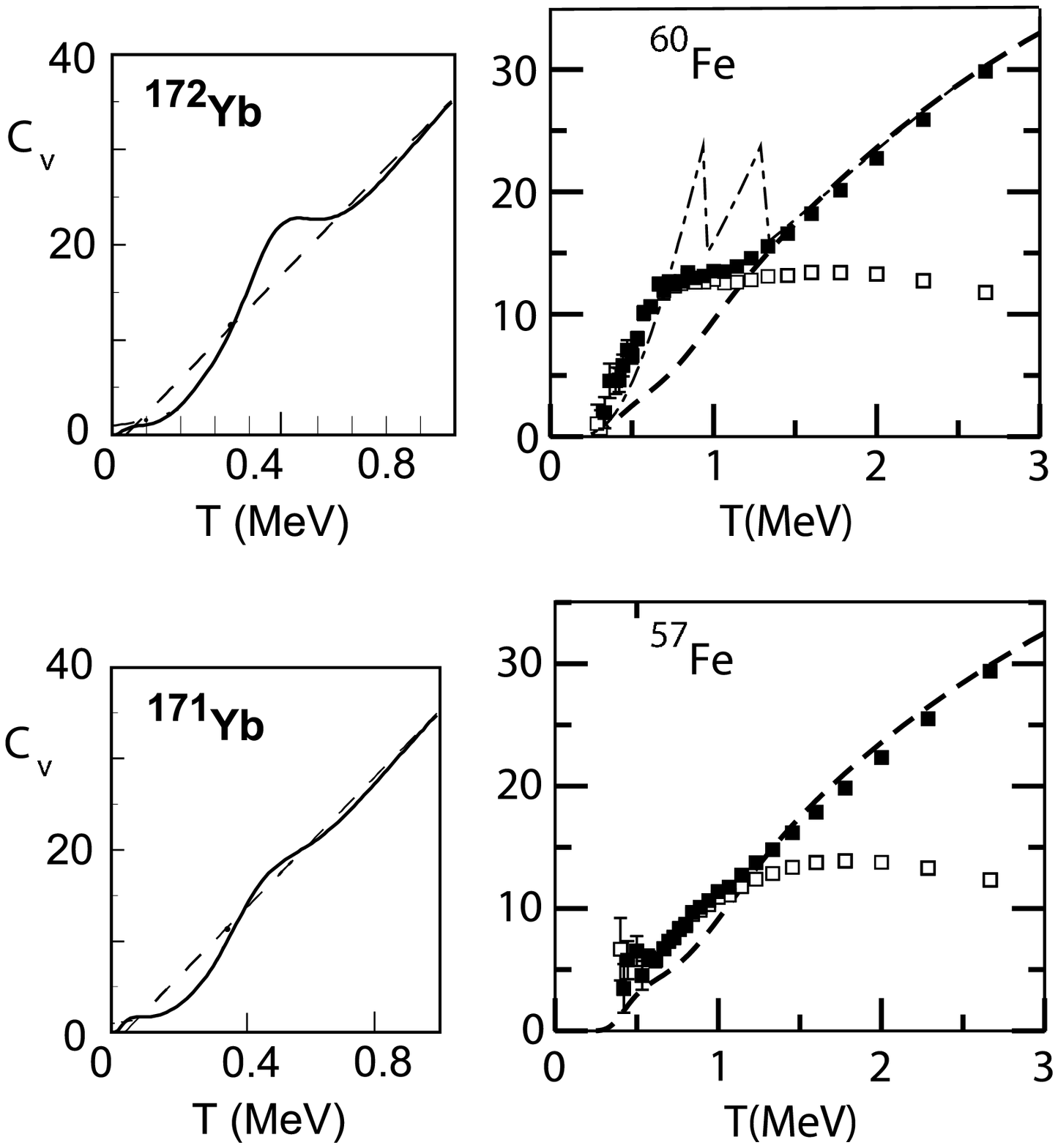}
\caption{Right panels: the heat capacities of $^{60}$Fe (top) and $^{57}$Fe (bottom). The extended heat capacity (solid squares) and the SMMC heat capacity in the $fpg_{9/2}$ shell (open squares) are compared with the heat capacity in the independent-particle model (dashed line). For $^{60}$Fe we also the BCS heat capacity (dotted-dashed). Left panels: the experimental heat capacities of  $^{172}$Yb and $^{171}$Yb (solid lines) in comparison with the Fermi gas results (dashed lines) \cite{Oslo}. Notice the similarity in the shape of the heat capacity between the theoretical and experimental results.} 
\end{figure}

 The pairing interaction leads to superconductivity in bulk metals below a critical temperature $T_c$. This phase transition is described by the BCS theory and predicts a discontinuity in the heat capacity at $T_c$ \cite{BCS}. The BCS theory is valid in the limit when the single-particle mean level spacing $d$ is much smaller than  the pairing gap $\Delta$. However, in finite nuclei, $d$ (within a shell) is comparable to $\Delta$, and fluctuations around the mean-field solution are important. An interesting question is whether signatures of the pairing transition can be still be observed despite the large fluctuations. 

  The heat capacity in SMMC is calculated as a numerical derivative of the thermal energy. Such calculations lead to large statistical errors at low temperatures. We have introduced a novel method that takes into account correlated errors and reduces the statistical error by almost an order of magnitude \cite{LA01}. Results for $^{60}$Fe and $^{57}$Fe are shown in the right panels Fig.~\ref{fig7}. The open squares are the truncated calculations (in the $fpg_{9/2}$ model space) \cite{LA01}, while the solid squares describe the extended heat capacity \cite{ABF03}. The heat capacity is suppressed in comparison with the BCS heat capacity (dotted-dashed line). However, in the even-even nucleus $^{60}$Fe we still observe a `bump' around $T \sim 0.8$ MeV when compared with the heat capacity of the 
independent-particle model (dashed line). This bump disappears almost entirely in the odd-even nucleus $^{57}$Fe and is a clear signature of the pairing transition.

  The heat capacity was recently measured in rare-earth nuclei \cite{Oslo}. Results (solid lines) are shown in the left panels of Fig.~\ref{fig7}) for the even-even nucleus $^{172}$Yb and the odd-even nucleus $^{171}$Yb.  These experimental heat capacities  have similar shapes to our calculated heat capacities.

\section{Conclusions}

  We have calculated statistical nuclear properties using the SMMC method. Statistical properties of particular importance in nuclear astrophysics are nuclear level densities and partition functions. In the Monte Carlo approach, fully correlated calculations are possible within complete major shells. We have extended our method to higher temperature by combining the correlated calculations in the truncated space with independent-particle model calculations in the full space, including the continuum. 

  Projected level densities at fixed values of the good quantum numbers can be calculated by incorporating exact projection techniques in the SSMC method. We have used such projection methods to calculate the parity and isospin distributions.

In the finite nucleus, fluctuations beyond
the mean field are important and smooth the singularities of the
phase transitions.  A particularly interesting example is the pairing transition. BCS theory predicts a discontinuity in the heat capacity at the transition temperature, but in the finite nucleus this signature is suppressed. Nevertheless, we do find a `bump' in the heat capacity of even-even nuclei around the pairing transition temperature, while such a signature is not seen in odd-even nuclei. We therefore conclude that in the fluctuation-dominated regime, thermal pairing correlations are manifested through strong odd-even effects. 


\begin{theacknowledgments}
  
  This work was supported in part by the Department of Energy grant No.\ DE-FG-0291-ER-40608. I would like to thank G.F. Bertsch, L. Fang, S. Liu and H. Nakada for their collaboration on various parts of the work presented above.
\end{theacknowledgments}


\end{document}